\documentclass[letterpaper,12pt]{article}
\usepackage{amsmath}
\usepackage{mathrsfs}
\usepackage{amsfonts}
\usepackage{amssymb}
\usepackage{graphicx}
\usepackage{physics}
\usepackage{latexsym}
\usepackage{epsfig}
\usepackage{pstricks}
\usepackage{stmaryrd}
\usepackage{rotating}
\usepackage[english]{babel}
\usepackage{setspace}
\usepackage[utf8]{inputenc}
\usepackage{natbib}
\usepackage[T1]{fontenc}
\usepackage{lmodern}
\usepackage{enumitem}
\usepackage{csquotes}
\usepackage{amsthm}
\usepackage{xcolor}
\usepackage{tikz}

\usepackage[normalem]{ulem} 
\usepackage[margin=1in]{geometry}
\begin{document}
\begin{center}	
\begin{LARGE}
\textbf{From locality to factorizability: a novel escape from Bell's theorem}\\
\end{LARGE}
\end{center}

\begin{center}
\begin{large}
G. S. Ciepielewski and E. Okon\\
\end{large}
 \textit{Universidad Nacional Aut\'onoma de M\'exico, Mexico City, Mexico.}\\[1cm]
\end{center}


While initial versions of Bell's theorem captured the notion of locality with the assumption of \emph{factorizability}, in later presentations, Bell argued that factorizability could be derived from the more fundamental \emph{principle of local causality}. Here we show that, contrary to what is commonly assumed, in order to derive factorizability from the principle of local causality, a non-trivial assumption, similar but strictly independent of settings independence, is required. Loosely speaking, such an extra assumption demands independence between the states of the measurement apparatuses. We conclude that it is possible to construct a model, satisfying both the principle of local causality and settings independence, but that, in virtue of violating this additional assumption---and thus factorizability---is able to break Bell's inequality. 


\onehalfspacing

\section{Introduction}

Through his famous theorem, John Bell challenged the capacity of local models to reproduce the predictions of quantum mechanics. In early versions of the theorem, Bell imposed the notion of locality with the assumption of \emph{factorizability}---in broad terms, the idea that correlations between distant systems must have local explanations. However, in later presentations, he argued that factorizability could be derived from his more general \emph{principle of local causality}. 

In this work, we show that, contrary to what is commonly accepted, to derive factorizability from the principle of local causality, an additional, non-trivial assumption is required. Loosely speaking, such an extra assumption demands independence between the states of the two measurement apparatuses. As such, the new assumption is similar, but strictly independent of \emph{settings independence}---an auxiliary assumption in Bell's theorem, demanding statistical independence between measurement settings and systems to be measured.

Given that, to derive factorizability from the principle of local causality, an extra assumption is required, we end up concluding that it is possible to construct a model, satisfying both the principle of local causality and settings independence, but which, in virtue of violating this additional assumption---and thus factorizability---can break Bell's inequality. Therefore, in light of experimental violations of the inequality, in the same way that one can retain locality by breaking settings independence (see \cite{ciepielewski2021}), one could retain both locality and settings independence by violating the new assumption; of course, as with violations of settings independence, the question of how \emph{reasonable} it is to violate the new assumption is independent of the purely logical point that it is possible to do so. 

In order to make our case, our manuscript is organized as follows. In section \ref{ss}, we review the standard story regarding Bell's theorem and the connection between the principle of local causality and factorizability. Then, in section \ref{lit}, we examine and criticize well-known discussions regarding such a connection. Next, in section \ref{rev}, containing the main result of this work, we show that in order to derive factorizability from the principle of local causality, an additional assumption is required. We also explore the nature of the new assumption and the possibility of violating it. Finally, in section \ref{con}, we offer our conclusions.

\section{The standard story} \label{ss}

Bell's construction of the theorem starts with an ensemble of pairs of particles in the singlet state. The particles of each pair are then sent to two spatially separated regions, 1 and 2, where spin measurements are performed. We denote by $a,b$ the spin directions measured in 1 and 2, respectively, and by $A,B$ the corresponding results (with spin-up corresponding to $+1$ and spin-down to $-1$). Next, Bell denotes by $\lambda$ the complete state of each pair. If quantum mechanics is complete, then $\lambda$ would be given by the quantum state; if quantum mechanics is not complete, then $\lambda$ could be the quantum state supplemented with extra variables or it could be something entirely different. Either way, Bell does not impose determinism on the models considered. 

Since Bell is interested in exploring the viability of local models, he assumes that the models under consideration have such a feature. To characterize them, he demands the probabilities $P\left(A,B|a,b,\lambda\right)$, predicted by such models for the experiment in question, to satisfy
\begin{equation}\label{fac}
P\left(A,B|a,b,\lambda\right) = P\left(A|a,\lambda\right) P\left(B|b,\lambda\right).
\end{equation}
This condition, usually referred to as \emph{factorizability}, is intended to capture the idea that, for local theories, all correlations between distant systems must have local explanations. That is, once one conditionalizes on the complete state $\lambda$ of a pair, correlations between distant measurements completely disappear.

Now, given that Bell considers measurements over an ensemble of pairs, such an ensemble can be characterized by a distribution over the fundamental states, $\rho(\lambda)$. The point is that, even though all pairs are prepared in the singlet state, the complete description given by $\lambda$ may change from pair to pair. Bell then assumes that the values of $\lambda$ and the settings $a$ and $b$ are statistically independent, that is
\begin{equation}\label{mi}
\rho(\lambda|a,b) = \rho(\lambda) .
\end{equation}
We call this condition \emph{settings independence}. Intuitively, it entails that the settings $a,b$ and $\lambda$ are not correlated. Settings independence seems reasonable because one can set up things in such a way that the settings can be chosen before, during, or after the generation of the pair, and they can be chosen using a variety of methods. Moreover, an analogous assumption is (implicitly) accepted in all experimental scenarios across all sciences.

To compose the theorem, Bell shows that, for any model satisfying factorizability (and settings independence), the expectation value of the product $AB$ over the whole ensemble,
\begin{equation} \label{expectation1}
E(a,b) = \int \sum_{A,B} A \, B \, P(A,B|a,b,\lambda) \, \rho(\lambda|a,b) \, d\lambda,
\end{equation}
necessarily obeys
\begin{equation}
|E(a,b) + E(a,b') + E(a',b) - E(a',b')| \leq 2 .
\end{equation}
That is, the predictions of all models satisfying factorizability (and settings independence), necessarily satisfy this inequality. 

Next, Bell considers the quantum prediction for the expectation value in Eq. (\ref{expectation1}), which is given by 
\begin{equation}
E^{QM}(a,b) = -\cos(\theta)
\end{equation}
with $\theta$ the angle between $a$ and $b$. Moreover, he notes that, if one takes $a,a',b,b'$ on the same plane, with a 90º angle between $a$ and $a'$ and $b$ and $b'$, and a 45º angle between $a$ and $b$, then
\begin{equation}\label{qprediction}
|E^{QM}(a,b) + E^{QM}(a,b') + E^{QM}(a',b) - E^{QM}(a',b')| = 2\sqrt{2}.
\end{equation}
It is clear, then, that quantum mechanics and models satisfying factorizability (and settings independence), make different predictions for the experiment in question.

The final step is to consider experimental realizations of Bell's scenario. Those experiments have been done and have established clear violations of the inequality \citep{aspect1981,aspect1982, weihs1998, giustina2015, shalm2015, hensen2015}. It seems, then, that models satisfying factorizability (and settings independence) are unable to correctly describe our world. These, in a nutshell, are Bell's theorem and its highly non-trivial implications.

Above we saw that the key condition employed by Bell to derive the inequality is factorizability. Initially, such a condition was assumed by Bell as a defining characteristic of local theories. However, in later presentations of the theorem, Bell argued that factorizability could be \emph{derived} from a more fundamental postulate, which he calls the \emph{principle of local causality} (see, e.g., \cite{bell1976,bell1990}).

According to Bell's principle of local causality, a model is local if the probability it assigns to $q_\chi$, the value of some physical quantity at spacetime event $\chi$, is such that
\begin{equation} \label{condprobA}
P(q_\chi|\lambda_\sigma) = P(q_\chi| \lambda_\sigma,q_\xi),
\end{equation} 
with $\lambda_\sigma$ a complete specification of the physical state on $\sigma$, a spatial slice of the past light cone of $\chi$, and $q_\xi$ the value of any physical quantity on $\xi$, an event spacelike separated from $\chi$ and outside of the causal future of $\sigma$ (see Figure 1).
\begin{figure}[ht]
\centering
\includegraphics[height=6cm]{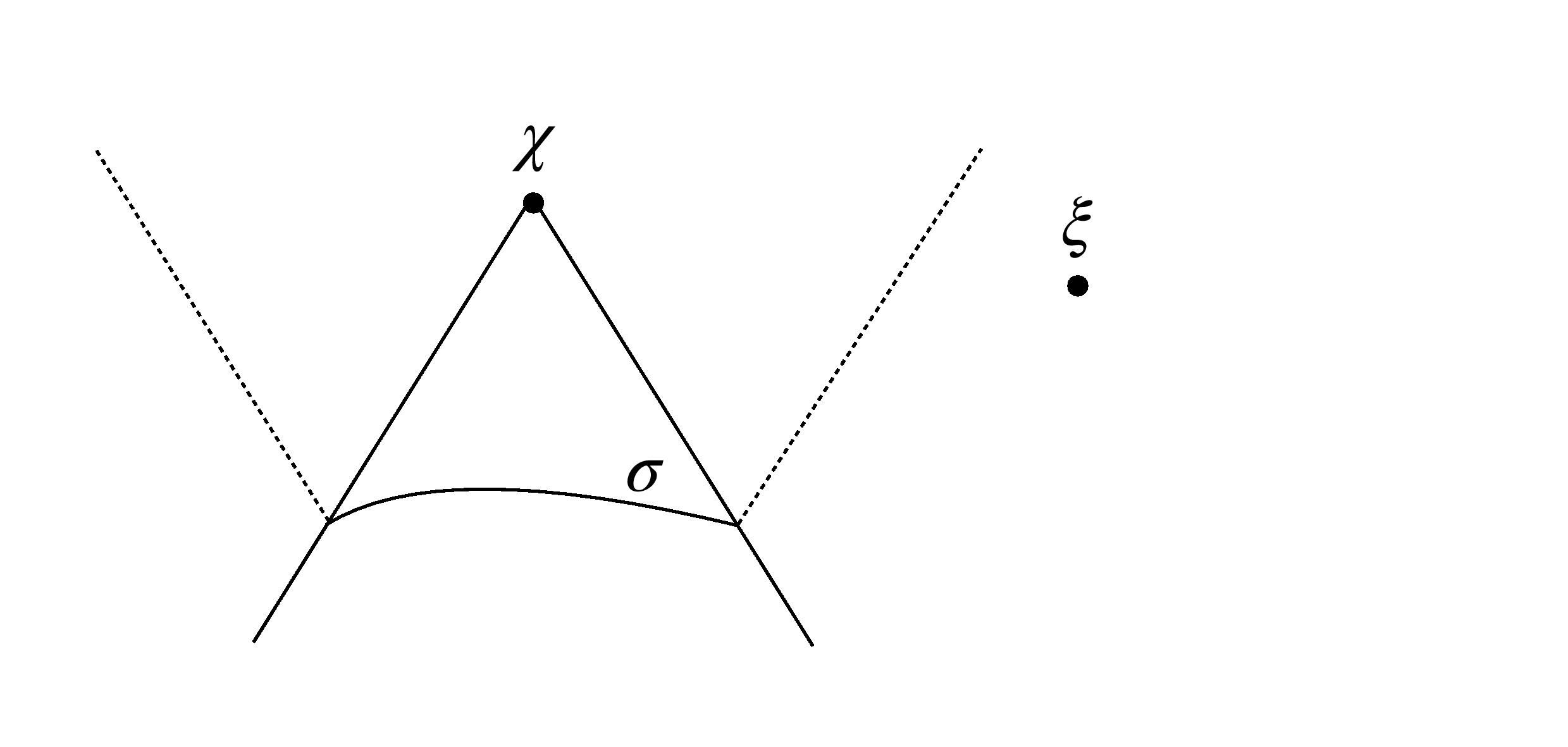} 
\caption{According to Bell's principle of local causality, a theory is local if $P(q_\chi|\lambda_\sigma) = P(q_\chi|\lambda_\sigma,q_\xi)$.}
\end{figure}
In words, the predictions that a local model makes for an event, given complete information on a slice of its past light cone, do not change with additional information about regions outside of the future of that slice.

Bell makes two important comments regarding his principle. First, it is crucial for $\xi$ to lie outside of the causal future of $\sigma$. If not, for a local, indeterministic theory, $\xi$ could contain information about a stochastic process to the future of $\sigma$, but in the common past of $\chi$ and $\xi$, that could improve the prediction of the model for $\chi$. Bell also stresses that it is essential for the specification of the physical state on $\sigma$ to be complete. Otherwise, even for local models, $\xi$ could contain part of the missing information, which could help improve the prediction.

As we said above, Bell claimed, and it is generally accepted (see, e.g., \cite{bell1976,bell1990,goldstein2011,maudlin2011,myrvold2021b,ciepielewski2021}), that factorizability follows from the principle of local causality. The connection is argued for along these lines. One considers the joint probability $P(A,B|a,b,\lambda_\Sigma)$, with $\lambda_\Sigma$ the complete state over $\Sigma$ (see Figure 2), and uses the product rule to write it as
\begin{equation}\label{PAB}
P(A,B|a,b,\lambda_\Sigma) = P(A|a,b,B,\lambda_\Sigma) P(B|a,b,\lambda_\Sigma) .
\end{equation}
\begin{figure}[ht]
\centering
\includegraphics[height=6cm]{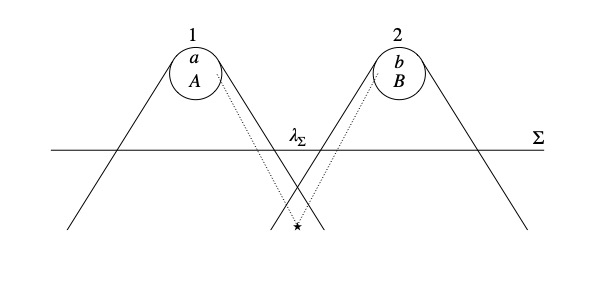} 
\caption{Space-time diagram of Bell's experimental scenario.}
\label{BES}
\end{figure}
Next, one employs the principle of local causality to note that
\begin{equation} 
P(A|a,b,B,\lambda_\Sigma) = P(A|a,\lambda_\Sigma) \quad \mbox{and} \quad P(B|a,b,\lambda_\Sigma) = P(B|b,\lambda_\Sigma) ,
\end{equation} 
from which it follows that 
\begin{equation} 
P(A,B|a,b,\lambda_\Sigma) = P(A|a,\lambda_\Sigma) P(B|b,\lambda_\Sigma).
\end{equation}
This looks like factorizability, but it isn't because, while factorizability is written in terms of $\lambda$, the complete state of the pair, the equation above is written in terms of $\lambda_\Sigma$, the complete state over $\Sigma$. To obtain factorizability, one then argues that $\lambda_\Sigma$ can be decomposed into two (not necessarily independent) parts: $\lambda$ and $\lambda_E$. As a result, Eq. (\ref{PAB}) is written as
\begin{equation} \label{LE}
P(A,B|a,b,\lambda, \lambda_E) = P(A|a,\lambda, \lambda_E) P(B|b,\lambda, \lambda_E).
\end{equation}
By explicitly having in the conditional $a$, $b$, $\lambda$ and $\lambda_E$, it is finally argued that one can remove $\lambda_E$ from the above equation. The idea is that, if one knows the complete state of the pair, $\lambda$, and one knows what is going to be measured, $a$ and $b$, then all additional information becomes irrelevant. With this, one finally arrives at
\begin{equation} 
P(A,B|a,b,\lambda) = P(A|a,\lambda) P(B|b,\lambda),
\end{equation}
that is, factorizability.

\section{A revision of the literature}
\label{lit}

Before delving, in the next section, into a detailed analysis of the connection between factorizability and the principle of local causality, in this section we examine a number of well-known discussions regarding the issue.

\subsection{La Nouvelle Cuisine}

We start this revision by exploring the use that Bell himself, in \emph{La Nouvelle Cuisine} \citep{bell1990}, gives to the principle of local causality for the derivation of the theorem. After stating the principle, and setting up the experimental scenario, Bell considers the joint probability of the results, $A$ and  $B$, conditional on four different terms: $a$, $b$, $\lambda$ and $c$; i.e., $P(A,B|a,b,c,\lambda)$. The first three terms on the conditional are, as expected, the settings and the state of the pair. Regarding $c$, Bell takes it to represent \emph{values of other variables describing the experiment}. However, he also assumes that $c$ and $\lambda$ give a specification which is \emph{complete}, at least for the intersection of $\Sigma$ with the union of the past light cones of regions 1 and 2 (see Figure 2). It seems, then, that counter to what Bell alleges, $c$ must contain much more than a description of the experiment and is much closer to what we called $\lambda_E$. That is, $c$ contains whatever is needed, besides $\lambda$, to make the description complete.

Next, Bell writes the joint probability he is considering as 
\begin{equation}
P(A,B|a,b,c,\lambda) = P(A|B,a,b,c,\lambda) P(B|a,b,c,\lambda),
\end{equation}
and, using the fact that the conjunction of $\lambda$ and $c$ is assumed to be complete, he employs the principle of local causality to arrive at
\begin{equation}
\label{PB}
P(A,B|a,b,c,\lambda) = P(A|a,c,\lambda) P(B|b,c,\lambda).
\end{equation}

At that point, and in contrast to what was done at the end of section \ref{ss}, Bell does not attempt to remove $c$ from the conditional. Instead, he keeps it in all conditionals throughout the derivation of the theorem. For instance, he writes the settings independence assumption as
\begin{equation}
\label{SIB}
\rho(\lambda|a,b,c) = \rho(\lambda|c).
\end{equation}
Moreover, he assumes that $c$ remains constant during the whole sequence of experiments. In fact, when he is done, he writes the inequality as
\begin{equation}
|E(a,b,c) - E(a,b',c)| + |E(a',b,c) + E(a',b',c)| \leq 2 .
\end{equation}

There is, however, a problem with retaining $c$ for the whole derivation and, in particular, with assuming it constant throughout the complete sequence of measurements. As we argued above, since the information provided by $c$ and $\lambda$ is assumed complete (which is what allowed for the principle of local causality to be used to arrive at Eq. (\ref{PB})), $c$ is required to contain much more information than the experimental arrangement. For instance, it must contain information regarding the temperature of the lab, the time of day or the material of the shoes of the experimentalists. It is fully unwarranted, then, to expect for $c$ not to change from run to run. We conclude, then, that Bell's derivation of the theorem, as presented in \emph{La Nouvelle Cuisine}, is simply invalid, as it depends on the invalid assumption that $c$ remains constant.

What about letting $c$ vary from run to run? That also leads to trouble. The issue is that, in that case, one would have to average over the distribution of $c$, call it $\mu(c)$, over the whole ensemble. However, given the definition of $c$, it is clear that such a distribution will \emph{not} be statistically independent of the settings $a$ and $b$, i..e, $\mu(c|a,b) \ne \mu(c)$. And, in the same way that, in the absence of settings independence, a possible dependence of $\lambda$ on the settings blocks the derivation of the inequality, a dependence of $c$ on $a$  and $b$ prevents the construction of the theorem (in fact, a similar averaging procedure was proposed in \cite{bell1976}, but criticized along the ideas of this paragraph in \cite{SHC}).

\subsection{Scholarpedia}

Next, we explore the discussion of the relation between factorizability and the principle of local causality in the popular Scholarpedia entry on Bell's theorem \citep{goldstein2011}. As we did in section \ref{ss}, (and translating everything into our notation), they start with the identity\footnote{Instead of using the surface $\Sigma$ to screen off, \citet{goldstein2011} employs a wider region (see their figure 2). Such a difference is not relevant for the following discussion.} 
\begin{equation}
P(A,B|a,b,\lambda_\Sigma) = P(A|a,b,B,\lambda_\Sigma) P(B|a,b,\lambda_\Sigma) 
\end{equation}
and use the principle of local causality to arrive at
\begin{equation}
\label{ScP}
P(A,B|a,b,\lambda_\Sigma) = P(A|a,\lambda_\Sigma) P(B|b,\lambda_\Sigma) .
\end{equation}
They readily acknowledge that such equality looks like factorizability, but it isn't. They recognize that $\lambda_\Sigma$ includes much more than $\lambda$ and that, while it is reasonable to assume that $\lambda$ is independent of the settings, it is not reasonable to assume that $\lambda_\Sigma$ is.

At this point, they claim that the principle of local causality alone is not sufficient to prove that there exists a $\lambda$, a subset of $\lambda_\Sigma$, which is independent of $a$ and $b$ and satisfies factorizability. The claim, in particular, is that the existence of this $\lambda$ also depends on the settings independence condition, which they call ``no conspiracy''. Moreover, they argue that by depending on the distinction between settings, controllable by experimenters, and other variables, the ``no conspiracy'' assumption involves anthropocentric elements. For this reason, they think, it is not possible to formalize the ``no conspiracy'' condition, nor to give a mathematical proof that, for a ``non-conspiratorial'' local theory, there exists a $\lambda$ which is independent of the settings and satisfies factorizability. Still, without any pretension of mathematical precision, they offer the following argument to the effect that, for a ``non-conspiratorial'' local theory, a $\lambda$ with such properties does exist.

The starting point of the argument is to distinguish between different subsets of $\lambda_\Sigma$. First, there are those values that are irrelevant for the experiment and can be ignored. Of those that are relevant, some, call them $\alpha$, will determine or influence the setting $a$ and some, call them $\beta$, the setting $b$. Finally, call whatever is left and relevant $\lambda_S$. Next, they notice that, for a ``non-conspiratorial'' theory, one must be able to define  $\alpha$,  $\beta$ and $\lambda_S$ such that $\lambda_S$ is independent of $a$ and $b$. Finally, to arrive at the claim that, if the theory is local, then factorizability must hold for this $\lambda_S$, they argue as follows. Since only $\lambda_S$ and $\alpha$ are relevant for $A$ and since $\alpha$ is relevant to $A$ only through $a$, then the motivation behind the principle of local causality leads to the conclusion that $P(A|a,b,B,\lambda_S) =  P(A|a,\lambda_S)$ (and analogously for region 2). If this is so, factorizability follows.

What are we to make of the discussion? We start by pointing out that, according to \citet{goldstein2011}, what needs to be proven is not factorizability alone, but ``the existence of a subset $\lambda$ of the data codified by [$\lambda_\Sigma$] that is independent of [$a,b$] and for which [factorizability] holds''. That is, from the unset, they formulate the issue in a way that factorizability and settings independence, or ``no-conspiracy'' as they call it, are linked. We believe that this approach only muddles the waters as factorizability and settings independence are two fully independent conditions. That is, it is perfectly possible to have either one without the other. For instance, the model in \cite{ciepielewski2021} satisfies factorizability but forbids the validity of settings independence, and pilot-wave theory violates factorizability but allows for settings independence to be the case.

Next, it is argued that settings independence involves anthropocentric elements, so it cannot be cleanly defined mathematically. We disagree. While it is clear that ``setting'' is an anthropocentric concept, that does not mean that the settings independence assumption cannot be defined precisely. As we mentioned when we introduced it, $\rho(\lambda)$ is the actual distribution of physical states $\lambda$ over the ensemble of measured pairs---and not a probability distribution. Moreover, settings independence consists of the assumption that such a distribution and the measurement settings, $a$ and $b$, are statistically independent: $\rho(\lambda|a,b) = \rho(\lambda)$. That is, one assumes that, if out of the whole ensemble of measured pairs, one focuses on a subensemble of runs with any particular pair of settings, then the distribution of $\lambda$ over that subensemble is the same as the distribution of the whole. It seems clear that, whether this is the case or not is a completely objective question, independent of the question of which variables are controllable by certain means (or even from whether there are controllable variables at all). That is, given a concrete experiment, and independently of how the settings were chosen, there will be an actual distribution of states and settings, and whether or not settings independence holds in that experiment is a completely objective question, with no anthropocentric elements rendering it subjective or vague.

It is important to add that the question of whether settings independence is satisfied or not is a question that can be asked of experiments, not models. Of course, there can be a models, such as the one in \cite{ciepielewski2021}, for which it is impossible to design a Bell-type experiment in which settings independence is the case. But that clearly does not mean that there are models for which settings independence is true for every experiment. For instance, if one is conducting a survey asking whether Real Madrid or Barcelona is the best football club in Spain, but conducts the survey in Barcelona, then the analog of settings independence will not be satisfied for that experiment---a fact that has little to do with any underlying model being able or not to satisfy settings independence.

Coming back to the discussion in \citet{goldstein2011}, we saw that, since factorizability and settings independence are blended, and since it is concluded that settings independence cannot be formalized, they settle for an informal argument. The starting point of that argument is the separation of $\lambda_\Sigma$ into four subsets: the irrelevant sector, $\alpha$ influencing $a$, $\beta$ influencing $b$, and the rest, which we called $\lambda_S$. After this, they argue that, for a ``non-conspiratorial'' model, $\lambda_S$ thus defined is independent of $a$ and $b$.

There are, however, a couple of complications with this reasoning. To begin with, as we just saw, $\lambda_S$ is not defined as the complete state of the pair to be measured, as with Bell, but as whatever is left after removing from $\lambda_\Sigma$ the irrelevant sector, $\alpha$, and $\beta$. We find this definition problematic for various reasons. First, by defining $\lambda_S$ this way, one also removes all the force behind the reasonableness of the settings independence assumption. That is, if $\lambda$ is the state of the pair to be measured, given a well-designed experiment, it is quite reasonable to assume that $\lambda$ and the settings are independent. This, however, is no longer the case for $\lambda_S$. It is perfectly possible for aspects of the measuring apparatuses, other than the settings, to be relevant for the result of an experiment (we will have much more to say about this in section \ref{rev}). If that is the case, the $\lambda_S$ defined above would include such aspects. But then it is clear that one would not be able to freely assume such features of the apparatuses, now part of $\lambda_S$, to be independent of the settings---i.e., one would not be able to justify the usage of settings independence. For instance, suppose that the result of the experiment depends on the $y$-position of a particular particle (so it is part of $\lambda_S$) and that the setting of one of the measuring devices depends on the $x$-position of the same particle (so it is part of, say, $\alpha$). If the particle is then constrained to be, say, in a circle, then $\lambda_S$ and the settings would clearly be correlated.

Of course, as pointed out in \citet{goldstein2011}, there are going to be models for which $\lambda_S$ will, in fact, be independent of the settings. We contend, however, that the important question to ask is not whether models like this exist, which they do, but whether, given a well-designed experiment, it is \emph{reasonable} to assume that $\lambda_S$ and the settings are independent. Our point is that this definition for $\lambda_S$ terribly weakens the force behind the reasonableness of this assumption.

Another potential problem with the definition of $\lambda_S$ is that it is experiment-dependent. That is, for the same system to be measured, the associated $\lambda_S$ would change when subjected to different measurements. This seems odd, as one would expect for $\lambda$ to capture aspects particular to the system to be measured, independently of aspects of the measurement apparatuses. Moreover, we find it problematic that $\lambda_S$, as defined in \citet{goldstein2011}, might not coincide with what a given model actually takes to be the state of the system to be measured.

There is, however, a bigger issue with the derivation of factorizability in \citet{goldstein2011}; an issue that prevails even after overlooking the problems with $\lambda_S$ mentioned above. The final step in their argument is the claim that, since only $\alpha$ and $\lambda$ are relevant for $A$, and since $\alpha$ is relevant to $A$ only through $a$, then $P(A|a,b,B,\lambda) =  P(A|a,\lambda)$. But why is it that $\alpha$ is relevant to $A$ only through $a$? It seems to us that there could be aspects of $\alpha$, beyond the information that $a$ will be the case, that could be useful to make a better prediction for $A$. Suppose, for instance, that the position of some particle influences, but not determines, the value of $a$ and determines the value for $A$. In that case, it is clear that the position of such a particle would be a part of $\alpha$, and it is so in a way that is relevant for $A$ beyond the fact that the setting is $a$. The point is that, in such cases, it is no longer true that one can substitute $\alpha$ for $a$ and $\beta$ for $b$, which would completely block the derivation of factorizability. That is, one would arrive at
\begin{equation}
P(A,B|\alpha,\beta,\lambda_S) = P(A|\alpha,\lambda_S) P(B|\beta,\lambda_S),
\end{equation}
but would not be able to derive factorizability from such an equation.

It could be argued that all aspects of $\alpha$, beyond the information that $a$ will be the case, and which could be useful to make a better prediction for $A$, would already be encoded in $\lambda_S$. But this is not so because, by definition, $\lambda_S$ is what remains \emph{after} removing $\alpha$ and $\beta$. Therefore, by definition, it does not contain anything belonging to $\alpha$. Moreover, even if one allows for those aspects of $\alpha$ to be contained in $\lambda_S$, that would only strengthen our case that this definition of $\lambda_S$ would make it unreasonable for settings independence to be the case.  

In sum, the discussion in \citet{goldstein2011} unnecessarily mixes the derivation of factorizability with the independent assumption of settings independence and introduces the non-standard $\lambda_S$, for which the reasons for assuming settings independence are removed. Moreover, the derivation of factorizability depends on the assumption that $\alpha$ is relevant to $A$ only through $a$ (and similarly for $\beta$, $b$ and $B$), but such an assumption is unwarranted---rendering the derivation invalid.


\subsection{Quantum Non-Locality and Relativity}

In order to justify factorizability, following \cite{bell1976}, chapter 4 of \emph{Quantum Non-Locality and Relativity} \citep{maudlin2011} calls $N$ the overlap of the past light cones of regions 1 and 2, $\Gamma$ the remainder of the past light cone of 1 and $M$ the remainder of the past light cone of 2 (see Figure 2). Moreover, it calls $\gamma$, $\mu$ and $\nu$ the complete specifications of the entire physical states of $\Gamma$, $M$ and $N$, respectively.\footnote{We use $\Gamma$ and $\gamma$, instead of Maudlin's $\Lambda$ and $\lambda$, to reserve $\lambda$ for the state of the pair, as defined in section \ref{ss}.} 

Next, two principles are invoked, Reichenbach's common cause principle and the principle of local causality. The first one is stated as entailing that if one takes into account all events that could play a causal role in bringing about a pair of events, then the correlations between them disappear. As for the second, it is taken to imply that all of the causes of an event must lie in its past light cone. Applying these principles to the particular case of Bell-type experiments leads to
\begin{equation}
\label{Pmau}
P(A,B|\gamma,\mu,\nu) = P(A|\gamma,\nu) P(B|\mu,\nu) .
\end{equation}

This, of course, is not factorizability. To arrive at such a condition, \cite{maudlin2011} argues that, when analyzing the Bell correlations, one usually stipulates that all of the causally relevant facts in region $N$ are contained in the state of the pair, which it's called $k$. Moreover, one assumes that the only event in $\Gamma$ which is relevant to the outcome is the setting $a$, and similarly for $M$. If so, $P(A|\gamma,\nu) = P(A|a,k)$ and $P(B|\mu,\nu) = P(B|b,k)$, so factorizability is obtained.

Are these assumptions legitimate? In footnote 10 of chapter 4 in \cite{maudlin2011}, it is argued that $k$ can, in fact, be stipulated to include all causally relevant physical facts in $N$. That might be so, the problem is that, as with $\lambda_S$ above, with this definition of $k$, all reasons to expect $k$ to be independent of the settings, i.e., all reasons to expect settings independence to be valid, disappear. Regarding the assumptions for the settings, things are even worse. In the same footnote, it is straightforwardly accepted that the settings simply cannot be stipulated to be the only causally relevant events in the remainders of the cones. As a possible remedy, the reader is then referred to Bell's treatment of the issue in \cite{bell1976}, in which he deals with other factors by averaging over them. The problem is that, as mentioned above, already in the year such a paper was published, \cite{SHC} pointed out that the derivation in \cite{bell1976} was, in fact, invalid.

\subsection{The Stanford Encyclopedia}

To conclude this assessment, we explore the article on Bell's Theorem in the The Stanford Encyclopedia of Philosophy \citep{myrvold2021b}. The discussion there starts by asserting that, in \cite{bell1976,bell1990}, factorizability is derived from the principle of local causality. Then, it goes on to distinguish between two versions of such a principle offered in \cite{bell1990}: the informal, referred to as PLC-1, which reads
\begin{quote}
The direct causes (and effects) of events are near by, and even the indirect causes (and effects) are no further away than permitted by the velocity of light,
\end{quote}
and the more formal, referred to as PLC-2, which is equivalent to the one we presented in section \ref{ss}. According to \citet{myrvold2021b}, PLC-2 follows from the conjunction of PLC-1 and Reichenbach's common cause principle.

Next, it is claimed that factorizability is the application of Bell's principle of local causality to the particular setting of Bell-type experiments. Moreover, after recalling the well-known fact \citep{jarrett1984} that factorizability can be derived from the conjunction of \emph{parameter independence} (PI) 
\begin{equation}
P(A|a,b,\lambda) = P(A|a,\lambda) \quad \text{and} \quad P(B|a,b,\lambda) = P(A|b,\lambda)
\end{equation}
and \emph{outcome independence} (OI)
\begin{equation}
P(A,B|a,b,\lambda) = P(A|a,b,\lambda) P(B|a,b,\lambda) ,
\end{equation}
the following blunt remark is offered
\begin{quote}
PI is a consequence of the causal locality condition PLC-1 alone, whereas OI requires in addition the assumption of the common cause principle.
\end{quote}
In sum, according to \citet{myrvold2021b}, the principle of local causality follows from PLC-1 and the common cause principle. Moreover, PI follows from PLC-1 and OI from PLC-1 and the common cause principle---and so from the principle of local causality. Putting everything together, factorizability, which follows from PI and OI, follows from the principle of local causality alone.

Are these claims correct? It is hard to say because no details about any of the derivations are given. Here, we will focus on the claims regarding PI and OI and show them to be problematic. We are also doubtful regarding the claim that the principle of local causality follows from PLC-1 and the common cause principle, but we will not explore that in detail.

Let's start with the claim that PLC-1 is sufficient for PI. To begin with, we point out that, while PLC-1 does not refer to probabilities, PI does. Therefore, one needs to add to PLC-1 a way of talking about probability. Granting this as unproblematic, the remaining task is with identifying as causes the settings $a,b$, and as effects the results $A,B$. If one could do this, PLC-1 would imply that the setting $b$ cannot affect the result $A$ and \emph{vice versa}. How, then, would one argue that the settings can be thought of as causes and the results as effects?

One possibility is by invoking an \emph{agency theory of causation}, according to which  ``the ordinary notions of cause and effect have a direct and essential connection with our ability to intervene in the world as agents'' \citep[p. 187]{menzies1993}. Since the settings $a$ and $b$ can, presumably,  be decided by the experimentalists, one can identify them as possible causes. If so, PLC-1 would imply that the setting $a$ ($b$) cannot affect the event $B$ ($A$). The Agency Theory of Causation, however, is not without its problems, \citep{woodward2003}. 

More serious complications besiege the derivation of OI from PLC-1 and the common cause principle---complications we have encountered before. To derive OI from the common cause principle, one would start by pointing out that, in a Bell-type experiment, there are correlations between the results $A$ and $B$, i.e.,  $P(A,B|a,b) > P(A|a,b) P(B|a,b)$. Then, by invoking the principle, there must be a common cause, $C$, such that, conditional on $C$, events $A$ and $B$ are independent. In other words, 
\begin{equation} \label{ccp}
P(A,B|a,b,C) = P(A|a,b,C) P(B|a,b,C).
\end{equation}
This surely looks like OI, but it isn't because $C$ may be different from $\lambda$: one is the Reichenbachian common cause and the other the complete state of the pair of particles---and it is perfectly possible for those two things not to be the same. One could then simply define $\lambda$ as $C$, but that is also no good because, in that case, as with previous attempts, the reasons we have for assuming $\lambda$ to be independent of the settings disappear.

More generally, there are two issues with attempts (see also \citet{cavalcanti2021}) at deriving factorizability from the principle of local causality and the common cause principle. The first, as we just saw, is that he have no control over the scope and nature of the common cause $C$, so it cannot be assumed to be $\lambda$, nor to be independent of the settings. The second issue is that, since, for instance, $a$ and $C$ cannot be assumed to give a complete physical description over a slice of the past light cone of region 1, one cannot apply the principle of local causality to remove $b$ from the first term in the right of the last equation, nor $a$ from the second. We conclude that, to properly derive factorizability, some other assumption must be included. We turn to that issue in the next section.

\section{From locality to factorizability revisited}
\label{rev}

In the previous section, we reviewed well-known attempts at deriving factorizability from the principle of local causality---and found them all wanting. In this section, we revisit the issue anew, looking for missing elements in past efforts. To do so, we start with the equality
\begin{equation}
\label{Pus}
P(A,B|a,b,\lambda) = P(A|B,a,b,\lambda) P(B|a,b,\lambda),
\end{equation}
with the standard definition for $\lambda$, i.e., that of the complete state of the pair to be measured. As discussed above, for such a $\lambda$ there are good reasons to assume, independently from factorizability and the principle of local causality, for it to be independent of the settings $a$ and $b$---i.e., there are good reasons to assume settings independence.

Now, to obtain factorizability from (\ref{Pus}) what we need is to remove $b$ and $B$ from the first term on the right, and $a$ from the second. Can we use the principle of local causality by itself to do so? Clearly not. Since $a$ and $\lambda$ do not constitute a complete description of a slice of the past light cone of the measurement in region 1, even though $b$ and $B$ are outside of such a cone, they cannot be simply removed, as they might contain information, not contained in $a$ and $\lambda$, useful for the prediction. Now, the fact that $b$ and $B$ \emph{might} contain such information does not imply that they do so. Therefore, the key question at this point is, under what circumstances would they? In other words, assuming the principle of local causality, under what circumstances would it be \emph{illegitimate} to remove $b$ or $B$ from the first term on the right-hand side of (\ref{Pus})?

For that to be the case, two conditions are required. First, there must be something, besides $a$ and $\lambda$, which is relevant for the prediction. Second, that something must be (at least partially) encoded in $b$ or $B$. Let's explore these two conditions with some care.

Regarding there being something besides $a$ and $\lambda$ relevant for the prediction, a straightforward possibility (already considered above), is for there being aspects of the measurement apparatus or its environment, other than the setting, determining or influencing the result. That is, there could be aspects of the full microscopic state of the apparatus and its environment in region 1, call it $\lambda_a$,\footnote{One could think of $\lambda_a$ ($\lambda_b$) as the complete physical state on a spacial slice of the past light cone of the apparatus on region 1 (2), and its immediate environment, moments before the measurement takes place.}  which clearly go beyond the information contained in $\lambda$ and $a$, that could be relevant for the result $A$. One concrete example in which this happens, adapted from a discussion of this possibility in \citet[p. 5]{spekkens2005}, is of a classical system and a classical measurement device, which generates an outcome by rolling one of several differently weighted dice with the choice of the die being determined both by the state of the system (call it $\lambda$) and the setting of the apparatus (call it $a$). It is quite clear, in this case, that the full microscopic state of the apparatus, (call it $\lambda_a$), would greatly improve upon the predictions one could make, conditional on $a$ and $\lambda$ alone: while the latter would limit to probabilistic statements, based on the die chosen, the former, being the underlying model deterministic, would allow for non-probabilistic, determined predictions.

We take a short detour to point out an important consequence of the previous discussion. It is often argued that any deterministic theory necessarily violates PI \citep{myrvold2016,butterfield2017}. This is important because PI seems to be incompatible with Lorentz invariance. Therefore, if determinism does imply violations of PI, the only chance for building a Lorentz invariant theory would be through indeterminism. The connection between determinism and violations of PI is argued for as follows. Any empirically adequate theory must violate Bell's inequality, so it must violate factorizability. Moreover, since factorizability can be derived from OI and PI, all such theories must violate one of these conditions. Next, it is argued that all deterministic theories trivially satisfy OI because, once the state and the settings are given, the probability of obtaining any result is either $1$ or $0$. Therefore, since OI must be the case in all deterministic theories, and because of the need to violate factorizability, it is concluded that any deterministic theory must violate PI. The problem with this argument, as we just saw, is that it is not the case that determinism implies satisfaction of OI, so it does not imply violations of PI. That is, even for a deterministic theory, it is not necessarily the case that the probability for $A$, once $\lambda$ and $a$ are given, must be either $1$ or $0$: the result may also depend on other aspects of $\lambda_a$. If all this is correct, it opens up the possibility for a deterministic, Lorentz invariant theory.

Coming back to the main discussion, what about the second condition, namely, for $\lambda_a$ to be, at least partially, encoded in $b$ or $B$? Well, as with settings independence, one must detach the purely logical question of its possibility with the more physical question of its reasonableness. Regarding the former, it is clear that the measurement apparatuses share a (not that distant) common past. Therefore, in the same way that it is possible for $\lambda$ to be correlated with the settings, it is perfectly possible for the microstate of the apparatus (and its environment) on one side to be correlated with the setting or result of the apparatus on the other. Moreover, it is clear that this condition is fully independent of settings independence: even if $\lambda$ and the settings are statistically independent, the microstate of the apparatus on one side could be correlated with the setting or result of the apparatus on the other.

Summing up, removing from the right-hand side of Eq. (\ref{Pus}) what is required to obtain factorizability is \emph{illegitimate} if i) besides the state and the settings, there are aspects of the apparatuses that are relevant for the prediction and ii) the microstate of the apparatus (or its environment) on one side has information regarding the setting or result on the other. Therefore, in order to enable the transition from locality to factorizability, at least one of these conditions must be blocked. What about denying the first condition, i.e., stipulating that only the state and the settings can be relevant for the prediction? The problem with this route is that it is hard to justify. As mentioned before, there are models for which additional variables, besides $a$ and $\lambda$, do influence $A$. Therefore, denying i) by fiat would amount to denying such models a priori. This seems antithetical to the whole project, manifested in Bell's work, of finding predictions that \emph{any local theory} would make. 

A better option is to deny the second condition. In particular, in order to enable the transition from locality to factorizability, it is sufficient to assume that the microstate of the apparatus and its and environment, on one side, are independent of the setting and result of the apparatus, on the other. That is, to demand
\begin{equation}
\label{MI}
P(\lambda_a|b,B) = P(\lambda_a) \quad \text{and} \quad P(\lambda_b|a,A) = P(\lambda_b),
\end{equation}
which we call (for lack of a better term) the \emph{microstate independence} assumption.\footnote{In fact, condition (\ref{MI}) is a little too strong, as all one actually needs in not independence between $b$ and $B$, and all of $\lambda_a$, but independence between $b$ and $B$, and the sector of $\lambda_a$ which is relevant for $A$, beyond $a$ and $\lambda$ (and analogously for the other side).} 

One may think that it would be simpler to enunciate the assumption as demanding independence between $\lambda_a$ and $\lambda_b$. That would be easier to state, and it would accomplish the same objective. The problem, though, is that independence between $\lambda_a$ and $\lambda_b$ is not a demand which is reasonable to make since, in general, the microstates of the measuring apparatuses will be correlated. For instance, if both are in the same lab, or even in the same city, we would expect correlations in their temperatures or in vibrations due to imperceptive earthquakes. Is the assumption, as stated above, more reasonable? We do think so, but we will explore the issue in due time.

With all this, we can enunciate the main result of this work: factorizability follows from the conjunction of the principle of local causality and the assumption of microstate independence. To see this, we start with Eq. (\ref{Pus}) and note that the principle of local causality implies that anything outside of the past light cone of 1, which does not enhance the prediction for $A$ given $a$ and $\lambda$, can be removed from the conditional in the probability of $A$ (and similarly for region 2). Then, we notice that, because of microstate independence, this is the case for $b$ and $B$ in the first term on the right and for $a$ in the second, so such terms can be removed, leading to factorizability. Now, since Bell's theorem follows from the conjunction of settings independence and factorizability, we conclude that Bell's theorem follows from the principle of local causality, settings independence and microstate independence. That, of course, means that it is possible to construct a model, satisfying both the principle of local causality and settings independence, which, in virtue of violating microstate independence---and thus factorizability---is able to break Bell's inequality.

To see how this would work in a concrete example, we can employ the game considered in \cite{maudlin2011}, in which two friends try to reproduce the quantum correlations displayed in Bell-type experiments. The rules of the game are as follows. The friends start in a room, where they can devise a strategy to coordinate their answers. Then, they leave by different doors and each is asked one of a list of possible questions, which they can answer with either ``up'' or ``down''. To enforce locality, it is assumed that, after leaving the room, the friends can no longer communicate. Of course, the point of the game is to show that, assuming settings independence---i.e., that there is no correlation between the strategy selected by the friends and the questions to be answered---the friends will not be able to reproduce the quantum correlations. 

What we want to show now is that, if one violates microstate independence, then the friends would be able to succeed in the game. To model a violation of microstate independence in this scenario, suppose that there is a correlation between the question to be asked on one side and, say, the color of the shirt of the examiner on the other (suppose, for concreteness, that the correlation is perfect). That is, in direct opposition to microstate independence, the setting (question) on one side is correlated with a feature of the apparatus (the color of the shirt of the examiner) on the other. It is quite clear that, under such circumstances, the friends would have no problem devising a strategy to reproduce the quantum statistics. For instance, by noting that the correlation between the color of the shirt on one side and the question on the other effectively means that one of the friends knows the question being asked to the other, the strategy could be to employ pilot-wave theory to produce the answers. That is, the friends could assume a singlet as the initial state and randomly choose in the room an initial condition analogous to the initial position of the pilot-wave particles. Then, they would choose the answers based on the corresponding predictions of pilot-wave theory for the analogous situation. In this way, all quantum predictions for the Bell experiment would be reproduced by a fully local procedure.

So far we have discussed the mere possibility of violating microstate independence, what about the reasonableness of doing so? Well, it could be argued that reasons very similar to those offered in support of settings independence, which we do find completing (although see \cite{ciepielewski2021}), would also be applicable here. It seems to us that this is indeed the case for the dependence between, say, $\lambda_a$ and $b$. That is, it seems to us that the same type of precautions usually considered to avoid a dependence between the settings and the state would also prevent, to a reasonable degree, dependencies between $\lambda_a$ and $b$. The independence between $\lambda_a$ and $B$, on the other hand, seems more tricky. The issue is that we seem to able to design experiments in which the settings are chosen in such a way that it is quite reasonable to assume for them to be independent of both the state and the microstate of the apparatus on the other side. And that seems to be helpful in preventing correlations between $\lambda_a$ and $b$. However, since we have no control over $B$, it is not that clear whether all these precautions would actually prevent $B$ from being correlated with $\lambda_a$. That is, while the reasons for assuming independence between $\lambda_a$ and $b$ seem strong, it is not clear that the same could be said regarding independence between $\lambda_a$ and $B$. 

Before moving on, it is useful to explore in some detail a possible dependence between the microstate on one side and the result on the other. First of all, it seems clear that a dependence between $\lambda_a$ and $B$ and between $\lambda_b$ and $A$, even in the absence of a dependence between $\lambda_a$ and $b$ and between $\lambda_b$ and $a$, would be sufficient for the inequality to be broken---as that would block the derivation of factorizability (note that a correlation between, say, $\lambda_a$ and $B$, in the absence of correlations between $\lambda_b$ and $A$, $\lambda_a$ and $b$, and $\lambda_b$ and $a$, would not block the derivation). Moreover, a model of that sort would display quite peculiar characteristics. Of course, it would share with standard quantum mechanics the ability to break OI. However, whereas in standard quantum mechanics $P(A|B,a,b,\lambda) \ne P(A|B,a,\lambda)$ and $P(A|B,a,\lambda) = P(A|a,\lambda)$, these models would satisfy the opposite, i.e., $P(A|B,a,b,\lambda) = P(A|B,a,\lambda)$ and $P(A|B,a,\lambda) \ne P(A|a,\lambda)$. That is, a model of this sort would manage to break IO through a very different path than standard quantum mechanics. We leave the analysis of models that employ these mechanisms to break the inequality for future work.

\section{Conclusions}
\label{con}

In early versions of Bell's theorem, the notion of locality was imposed via the factorizability assumption. However, in later presentations, Bell introduced the more general and fundamental principle of local causality and argued that factorizability could be derived from such a principle. In this work, we have analyzed prominent attempts at such a derivation and have shown that, contrary to what is usually affirmed, the derivation requires the introduction of an additional, non-trivial assumption---which we call microstate independence. This new assumption is similar to, but strictly independent of the well-known settings independence assumption, and demands independence between the microstate of the apparatus on one side and the setting and result of the apparatus on the other.

Given the fact that, to derive factorizability from the principle of local causality, this additional assumption is required, we conclude that it is possible to construct a model that satisfies settings independence and the principle of local causality, but which, in virtue of breaching this new assumption---and thus factorizability---is able to break Bell's inequality. In fact, we display a toy model in which this is the case. We close by pointing out that, in view of clear experimental violations of Bell's inequality, in the same way that one can cling to locality by giving up settings independence, one could hold on to, both, locality and settings independence, by dropping the microstate independence assumption.

Regarding the reasonableness of doing so, we maintain that, for the independence between microstates on one side and settings on the other, arguments very similar to those usually provided in support of settings independence, which we do find (for the most part) completing, would also apply in this case. Such arguments might be considered less effective in defending the independence between microstates on one side and results on the other, as we have no control over results. Still, it seems to us that correlations between microstates on one side and results on the other would strike most as quite conspiratorial; we leave further discussion of this point for the future.

\section*{Acknowledgments}

We thank Shelly Goldstein for valuable comments. We acknowledge support from CONACYT grant 140630 and PAPIIT grant IN405320.

\bibliographystyle{apalike}
\bibliography{factorizability}
\end{document}